\documentclass[twocolumn,amsmath,amssymb,osajnl]{revtex4}
\usepackage{epsfig}

\begin{document}

\title{Modeling of realistic cladding structures for air-core photonic band-gap fibers}

\author{Niels Asger Mortensen}
\affiliation{Crystal Fibre A/S, Blokken 84, DK-3460 Birker\o d, Denmark}

\author{Martin Dybendal Nielsen}
\affiliation{Crystal Fibre A/S, Blokken 84, DK-3460 Birker\o d, Denmark\\COM, Technical University of Denmark, DK-2800 Kongens Lyngby, Denmark}

\begin{abstract}
Cladding structures of photonic band-gap fibers often have air-holes of non-circular shape and, typically, close-to-hexagonal air holes with curved corners are observed. We study photonic band-gaps in such structures by aid of a two-parameter representation of the size and curvature. For the fundamental band-gap we find that the band-gap edges (the intersections with the air line) shift toward shorter wavelengths when the air-filling fraction $f$ is increased. The band-gap also broadens and the relative band-width increases exponentially with $f^2$. Comparing to recent experiments [Nature {\bf 424}, 657 (2003)] we find very good agreement.
\end{abstract}


\maketitle

In air-silica photonic crystal fibers (PCFs) an arrangement of air-holes running along the full length of the fiber provides the confinement and guidance of light in a defect region. For photonic band-gap (PBG) guiding PCFs the air-holes have been arranged in various ways such as in a triangular lattice\cite{cregan1999}, but honey-comb\cite{knight1998} and kagome\cite{benabid2002} arrangements are other options. Cregan {\it et al.}\cite{cregan1999} have demonstrated that light can be guided in an air-core by means of the PBG effect and this observation has stimulated an avalanche of both basic and applied research. For recent reviews we refer to Refs.~\onlinecite{russell2003,knight2003} and references therein.

\begin{figure}[b!]
\begin{center}
\epsfig{file=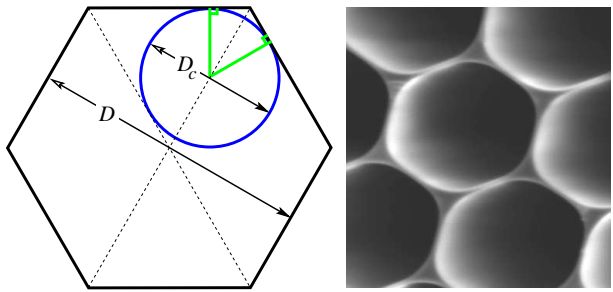, width=0.45\textwidth,clip}
\end{center}
\caption{The right panel shows a scanning-electron micrograph of the cladding structure of a PBG fiber with a pitch of $\Lambda\simeq 2.7\,{\rm\mu m}$. The left panel illustrates the two-parameter representation of the air-hole shape. }
\label{fig1}
\end{figure}

From a modeling point of view the air-holes are generally assumed circular\cite{birks1995,broeng2000} and for fibers fabricated by the stack-and-pull method this is also typically what is observed for moderate air-filling fractions.\cite{knight1996} However, for the high air-filling fractions $f>0.8$ recently employed for PBG guidance the air-holes tend to be non-circular.\cite{smith2003,bouwmans2003} From a simple geometric consideration it is easily shown that the theoretical upper limit of the air-filling fraction in a structure of close packed circular voids is $f=\pi/(2\sqrt{3})\simeq 0.91$. Hence, design and fabrication of air-core PBG fibers call for a detailed modeling of the spectral position of the PBGs which goes beyond the assumption of circular air holes. In this Letter we consider the triangular cladding arrangement (with a pitch $\Lambda$) first employed by Cregan {\it et al.},\cite{cregan1999} but for the air-holes we take the non-circular shape observed recently in Refs.~\onlinecite{smith2003,bouwmans2003} into account. From scanning-electron micrographs (SEMs) we find that the air-holes are of over-all hexagonal shape with curved corners, see Fig.~\ref{fig1} (this conclusion was also emphasized in Ref.~\onlinecite{smith2003}). This suggests that the holes can be parametrized by two parameters; the edge-to-edge distance $D$ (corresponding to the diameter of a circular air-hole) and the diameter of curvature, $D_c$, at the corners (see left panel of Fig.~\ref{fig1}). In this two-parameter representation the air-filling fraction is given by

\begin{eqnarray}
f&=&\frac{A_{\rm hex}(D)-A_{\rm hex}(D_c)+A_{\rm circ}(D_c)}{A_{\rm hex}(\Lambda)}\nonumber\\
&=&\bigg(\frac{D}{\Lambda}\bigg)^2\bigg[1 - \bigg(1-\frac{\pi}{2\sqrt{3}}\bigg)\bigg(\frac{D_c}{D}\bigg)^2\bigg]\label{f}
\end{eqnarray}
where $A_{\rm hex}(x)=\sqrt{3}\,x^2/2$ and $A_{\rm circ}(x)=\pi (x/2)^2$ are the areas of a hexagon (with edge-to-edge distance x) and a circle (of diameter $x$), respectively. 

For the optical properties we apply Maxwell's equations to a macroscopic and isotropic loss-less dielectric medium and assume a linear dependence of the displacement field on the electrical field. We consider a harmonic mode ${\boldsymbol H}({\boldsymbol r},t)={\boldsymbol H}_\omega({\boldsymbol r})e^{i\omega t}$ with angular frequency $\omega$ and substituting into Maxwell's equations the magnetic-field vector is then governed by the wave equation\cite{joannopoulos}
\begin{equation}\label{waveequation}
{\boldsymbol \nabla}\times \frac{1}{\varepsilon({\boldsymbol r})} {\boldsymbol \nabla}\times {\boldsymbol H}_\omega({\boldsymbol r})=\frac{\omega^2}{c^2}{\boldsymbol H}_\omega({\boldsymbol r}).
\end{equation}
Here, $c$ is the velocity of light and $\varepsilon$ is the dielectric function which we in the following assume independent of frequency by which the wave equation becomes scale-invariant. All the results to be presented can thus be scaled to the desired value of $\Lambda$.

\begin{figure}[t!]
\begin{center}
\epsfig{file=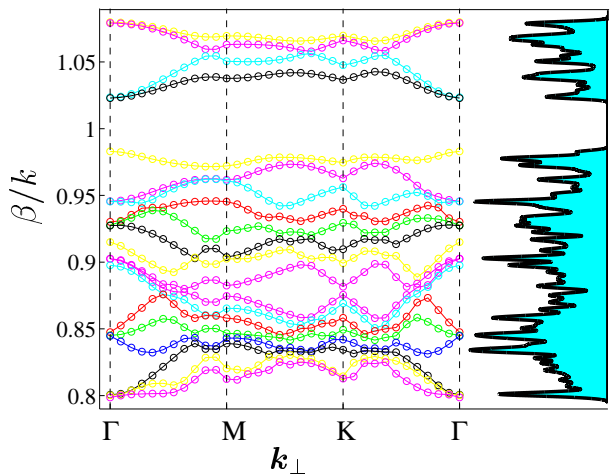, width=0.45\textwidth}
\end{center}
\caption{Plot of $\beta/k$ versus ${\boldsymbol k}_\perp$ along principal directions in the Brillouin zone for $\beta\Lambda=14.45$ and a structure with $D/\Lambda=0.96$ and $D_c/\Lambda=0.55$ (see inset of Fig.~\ref{fig3}). The filled curve shows the corresponding density-of-states (the projection of the data onto the $y$-axis). }
\label{fig2}
\end{figure}

For a PCF geometry, the cladding dielectric function $\varepsilon({\boldsymbol r})$ is periodic in the transverse plane and translational invariant along the fiber-axis (the $z$-axis). The solution is according to Bloch's theorem a plane wave modulated by a function ${\boldsymbol h}_\omega({\boldsymbol r}_\perp)$ with the periodicity of the dielectric structure in the transverse direction
\begin{equation}
{\boldsymbol H}_\omega({\boldsymbol r})={\boldsymbol h}_\omega({\boldsymbol r}_\perp)\exp(i{\boldsymbol k}_\perp\cdot {\boldsymbol r}_\perp + i\beta z).
\end{equation}
Substituting this ansatz into Eq.(\ref{waveequation}) we get an eigenvalue problem for $\omega({\boldsymbol k}_\perp,\beta)$ which we solve by the aid of a plane-wave basis (typically $128\times128$ plane waves) with periodic boundary conditions.\cite{johnson2001} For the dielectric function we use $\varepsilon=1$ in air and $\varepsilon=(1.444)^2=2.085$ in silica. From a computational point of view our simulations thus follow the same lines as those used for structures with circular holes.\cite{broeng2000}

\begin{figure}[b!]
\begin{center}
\epsfig{file=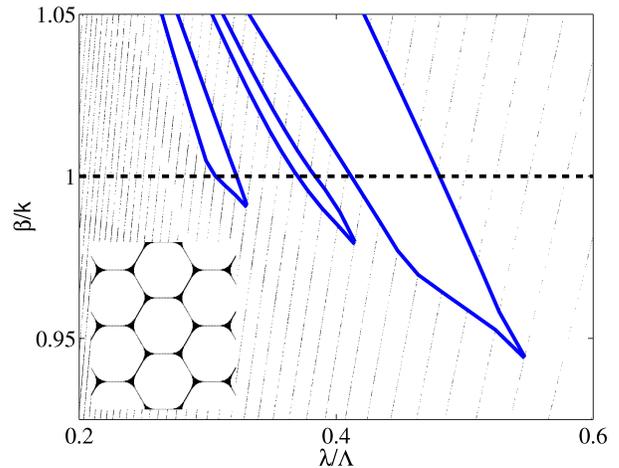, width=0.45\textwidth,clip}
\end{center}
\caption{Plot of $\beta/k$ versus $\lambda/\Lambda$ for a structure with $D/\Lambda=0.96$ and $D_c/\Lambda=0.55$ (see inset). The data-points result from a sampling of the Brillouin zone along the usual path from the $\Gamma$-point to the $M$-point and via the $K$-point back to the $\Gamma$-point. The solid lines indicate the band-gap boundaries and the dashed line $\beta/k=1$ shows the air line which passes through several band-gaps. }
\label{fig3}
\end{figure}

In Fig.~\ref{fig2} we show photonic bands calculated for $\beta\Lambda=14.45$ and a structure with $D/\Lambda=0.96$ and $D_c/\Lambda=0.55$ (see inset of Fig.~\ref{fig3}). It is common to introduce the free-space wave-number $k=2\pi/\lambda=\omega/c$ and plot the effective index $\beta/k=c\beta/\omega$ rather than the angular frequency $\omega$. The bands are plotted as a function of ${\boldsymbol k}_\perp$ along the usual principal directions in the Brillouin zone; from the $\Gamma$-point to the $M$-point and via the $K$-point back to the $\Gamma$-point.\cite{broeng2000} As seen the structure exhibits a band-gap which is particular clear from the filled curve which shows the corresponding density-of-states (the projection of the data onto the $y$-axis). In this example the band-gap is around the index of air ($\beta/k=1$), but in general the position and width of the band-gaps depend on the propagation constant $\beta$. In Fig.~\ref{fig3} we show the results of changing $\beta$.  The data-points result from a sampling of the Brillouin zone along the above mentioned principal directions. The PBGs are those regions with no data-points and the solid lines indicate the corresponding band-gap edges. The dashed line shows the air-line which passes through several band-gaps. Usually the band-gap at the longest wavelength (the lowest frequency) is refereed to as the fundamental band-gap and the other gaps are denoted higher-order band-gaps.\cite{broeng2000} The slopes of the band-gaps are relatively large which suggest that e.g. scattering and bending-loss will increase almost abruptly when the wavelength approaches the band-edges.

\begin{figure}[t!]
\begin{center}
\epsfig{file=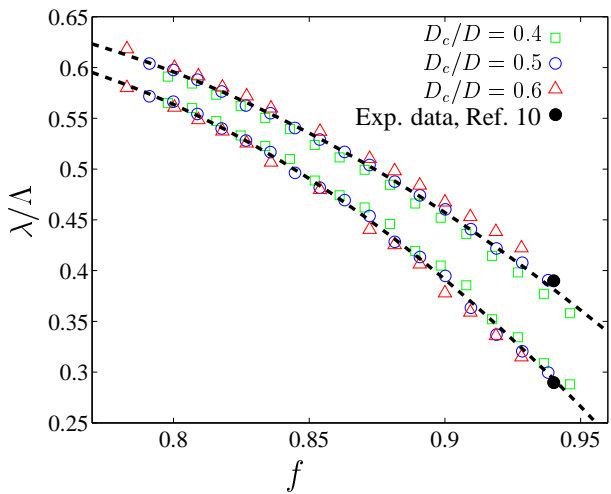, width=0.45\textwidth,clip}
\end{center}
\caption{Plot of the band-gap edges $\lambda_u$ and $\lambda_l$ of the fundamental band-gap as a function of air-filling fraction $f$. The dashed lines are guides to the eyes. Recent experimental data\cite{smith2003} is also included.}
\label{fig4}
\end{figure}

For PBG fibers with an air-core\cite{cregan1999,smith2003,bouwmans2003} the band-edges of guided modes will to a good approximation be given by the intersection of the air-line with the cladding band-edges in Fig.~\ref{fig3}. For the fundamental band-gap we denote the upper and lower intersections by  $\lambda_u$ and $\lambda_l$, respectively. We have calculated the band edges for various values of  $D$ and $D_c$; Fig.~\ref{fig4} summarizes the results. The band-diagrams of course depend on the two parameters $D$ and $D_c$, but we find that a single-parameter description of the band-edges is possible in terms of the air-filling fraction $f$, Eq.~(\ref{f}).

\begin{figure}[t!]
\begin{center}
\epsfig{file=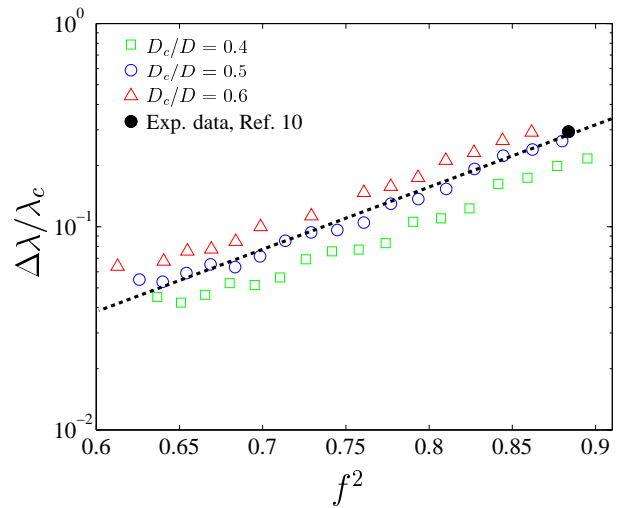, width=0.45\textwidth,clip}
\end{center}
\caption{Relative band-width versus $f^2$. The straight dashed line indicates the close-to-exponential dependence, Eq.~(\ref{dlambda}). Recent experimental data\cite{smith2003} is also included.}
\label{fig5}
\end{figure}

When the air-filling fraction is increased the center $\lambda_c=(\lambda_u+\lambda_l)/2$ of the band-gap shifts toward shorter wavelengths. Furthermore, the band-width $\Delta\lambda=\lambda_u-\lambda_l$ at the same time increases and the relative band-width shows a close to exponential increase (see Fig.~\ref{fig5}), {\it i.e.}

\begin{equation}\label{dlambda}
\Delta\lambda/\lambda_c\sim {\mathcal B} \times \exp({\mathcal A} \times f^2),
\end{equation}
where $\mathcal A$ and $\mathcal B$ are positive numerical coefficients. The results in Figs.~\ref{fig4},\ref{fig5} support the choice of high air-filling fractions for practical fibers and as seen we find very good agreement between our numerics and recent experimental results by Smith {\it et al.}\cite{smith2003}

In summary, we have shown that realistic PBG cladding structures\cite{smith2003,bouwmans2003} can be represented by an ``ideal'' two-parameter description which facilitate detailed numerical modeling. For the fundamental band-gap the band-gap edges (the intersections with the air line) shift toward shorter wavelengths for an increasing air-filling fraction $f$ and the band-gap also broadens significantly. This observation may make air-core PBG fibers realistic for wavelengths even shorter than the 850 nm reported recently by Bouwmans {\it et al.}\cite{bouwmans2003}

\vspace{5mm} 
We are grateful to B.~H. Larsen (NKT Research) for providing SEMs of real samples and our colleges T.~P. Hansen, J. Broeng, and G. Vienne  and for stimulating discussions on the experimentally observed structures. M.~D. Nielsen acknowledges financial support by the Danish Academy of Technical Sciences. N.~A. Mortensen's e-mail address is asger@mailaps.org.

\end{document}